\documentclass{aastex62}
\usepackage{graphics}

\submitjournal{ApJS}

\begin{document}

\title{An XMM--Newton Early-type Galaxy Atlas}

\author{Nazma Islam}
\altaffiliation{Present affiliation: 1. Center for Space Science and Technology, University of Maryland,
Baltimore County, 1000 Hilltop Circle, Baltimore, MD 21250, USA \\
2. X-ray Astrophysics Laboratory, NASA Goddard Space Flight Center,
Greenbelt, MD 20771, USA}
\affil{Center for Astrophysics $\mid$ Harvard $\&$ Smithsonian, 60 Garden Street, Cambridge, MA 02138, USA}

\correspondingauthor{Nazma Islam}
\email{nislam@umbc.edu}

\author{Dong-Woo Kim}
\affil{Center for Astrophysics $\mid$ Harvard $\&$ Smithsonian, 60 Garden Street, Cambridge, MA 02138, USA}
\correspondingauthor{D.-W. Kim}
\email{dkim@cfa.harvard.edu}

\author{Kenneth Lin}
\affil{Department of Astronomy, University of California, Berkeley, CA 94720-3411, USA}
\affil{Lawrence Berkeley National Laboratory, 1 Cyclotron Road, Berkeley, CA 94720, USA}

\author{Ewan O'Sullivan}
\affil{Center for Astrophysics $\mid$ Harvard $\&$ Smithsonian, 60 Garden Street, Cambridge, MA 02138, USA}

\author{Craig Anderson}
\affil{Center for Astrophysics $\mid$ Harvard $\&$ Smithsonian, 60 Garden Street, Cambridge, MA 02138, USA}

\author{Giuseppina Fabbiano}
\affil{Center for Astrophysics $\mid$ Harvard $\&$ Smithsonian, 60 Garden Street, Cambridge, MA 02138, USA}

\author{Jennifer Lauer}
\affil{Center for Astrophysics $\mid$ Harvard $\&$ Smithsonian, 60 Garden Street, Cambridge, MA 02138, USA}

\author{Douglas Morgan}
\affil{Center for Astrophysics $\mid$ Harvard $\&$ Smithsonian, 60 Garden Street, Cambridge, MA 02138, USA}

\author{Amy Mossman}
\affil{Center for Astrophysics $\mid$ Harvard $\&$ Smithsonian, 60 Garden Street, Cambridge, MA 02138, USA}

\author{Alessandro Paggi}
\affil{INAF- Osservatorio Astrofisico di Torino, via Osservatorio 20, 10025 Pino Torinese, Italy}

\author{Ginevra Trinchieri}
\affil{INAF-Osservatorio Astronomico di Brera, via Brera 28, 20121 Milano, Italy}

\author{Saeqa Vrtilek}
\affil{Center for Astrophysics $\mid$ Harvard $\&$ Smithsonian, 60 Garden Street, Cambridge, MA 02138, USA}

\begin{abstract}
The distribution of hot interstellar medium in early-type galaxies bears the imprint of the various astrophysical processes it underwent during its evolution. The X-ray observations of these galaxies have identified various structural features related to AGN and stellar feedback and environmental effects such as merging and sloshing. In our XMM-Newton Galaxy Atlas (NGA) project, we analyze archival observations of 38 ETGs, utilizing the high sensitivity and large field of view of XMM-Newton to construct spatially resolved 2D spectral maps of the hot gas halos. To illustrate our NGA data products in conjunction with the Chandra Galaxy Atlas \citep{kim2019}, we describe two distinct galaxies -- NGC 4636 and NGC 1550, in detail. We discuss their evolutionary history with a particular focus on the asymmetric distribution of metal-enriched, low-entropy gas caused by sloshing and AGN-driven uplift.  We will release the NGA data products to a dedicated website, which users can download to perform further analyses. 

\end{abstract}


\section{Introduction}
\label{intro}

The hot interstellar medium (ISM) in Early-Type Galaxies (ETGs) plays an important role in understanding their formation and evolutionary history. Various astrophysical processes such as AGN feedback, stellar feedback, environmental effects such as  merging, sloshing, tidal stripping, etc., leave an imprint on the distribution of the hot gas in these ETGs (e.g., see \citealt{kim2012} and references therein). Although the optical images of these galaxies show a smooth and featureless distribution, the X-ray surface brightness images may reveal the presence of asymmetry in the distribution of the hot gas in these galaxies. Two decades of observations with Chandra and XMM-Newton telescopes have revolutionised our understanding of the distribution of hot gas in ETGs. The unprecedented spatial resolution of Chandra and high sensitivity and large field of view of XMM have allowed us to study in detail the various asymmetric distributions in the hot gas like cold fronts, bubbles, filaments and X-ray tails which are indicative of the different astrophysical processes remnant or ongoing in the ETGs.
\par
Previous archival studies on large samples of ETGs focussed mainly on the 1D radial profiles and global properties like scaling relations etc \citep{osullivan2003, diehl2007, diehl2008a, diehl2008b, kim2013, kim2015, babyk2018, lakhchaura2018, goulding2016}. 
As part of the Chandra Early-type Galaxy Atlas project (CGA), \citep[hereafter K19 in the paper]{kim2019} systematically analysed Chandra observations of 70 E and S0 type galaxies, with the objective of creating spatially resolved 2D intensity and spectral maps, as well as 1D radial profiles. 
These 2D spectral maps are important in revealing unique features in the distribution of the hot gas, which might not be discernible in 1D radial profiles or the 2D surface brightness maps. CGA\footnote{http://cxc.cfa.harvard.edu/GalaxyAtlas/v1/} utilises  robust data analysis pipelines, especially four different spatial binning techniques, to uniformly analyse a large dataset of ETGs. 
\par
In this paper, we utilise the large field of view and higher sensitivity of XMM to systematically carry out uniform data analysis of 38 ETGs. The large field of view of XMM-Newton is crucial in studying the  diffuse gas emission in the outskirts of the galaxies by measuring their spectral properties and mass profile on a larger scale. This is critical for understanding the interaction of this hot gas with the surrounding medium (e.g., by ram pressure stripping) and neighbouring galaxies (e.g., sloshing, merging). The larger effective area of XMM compared to Chandra is also important in studying the metal abundances, especially Fe abundances in the ETGs. Metal abundances in the hot gas of ETGs are the relics of stellar and chemical evolution. They are related to the stellar mass loss rate and supernova ejecta, hence provide important information about the metal enrichment history of the hot gas in ETGs \citep{ji2009, kim2012b, panagoulia2015}. In this paper, we present Fe abundance maps which were not included in CGA.
\par
The paper is organised in the following order: Section 2 describes the sample selection and the XMM observations. Section 3 describes the data analysis techniques, especially the robust pipelines developed for this project. In Section 4, we describe the results for two galaxies, NGC 4636 and NGC 1550, and their implications on understanding the various astrophysical processes that have affected them. Throughout the paper, the errors are quoted at 1$\sigma$ level.   

\section{Sample Selection and XMM-Newton observations} 
\label{sample}
Our sample consists of 38 E and S0 galaxies, similar sample used in K19 and for which observations are available from the public XMM archive \footnote{https://www.cosmos.esa.int/web/xmm-newton/xsa}. XMM carries three X-ray imaging instruments, two EPIC-MOS cameras and the EPIC-PN cameras. To avoid cross-calibration issues while using these two different types of CCDs and uniformly analyse a large set of observations, we only use observations with MOS cameras (MOS 1 and MOS 2). 
\par
Table 1 tabulates the sample of ETGs observed with XMM and analysed as part of NGA project, with basic galaxy information (co-ordinates, distance, galaxy type, size, and K band luminosities) along with the effective exposures in MOS after removing the background flares. Some of the galaxies have been observed more than once by XMM. We have devised a nomenclature for our sample, Merge ID (mid): we assign a five digit number following the galaxy name, the first digit is 9, followed by mm (where mm is the number of useful observations with XMM; 01 for single observations, 02 for two observations etc), and the last two digits (usually ``01") are unique serial numbers to separate different combinations of pipeline parameters (e.g., in background flare, point source detection, source regions).  
\par
Table 2 lists the XMM ObsIDs used for each galaxy included in the analysis, the observation dates, the off-axis angle (OAA; in arcmin) of the galaxy center from the telescope aim point and the effective exposures for each observation after filtering for background flares.

\begin{figure*}
\centering
 \includegraphics[scale=0.4,angle=0]{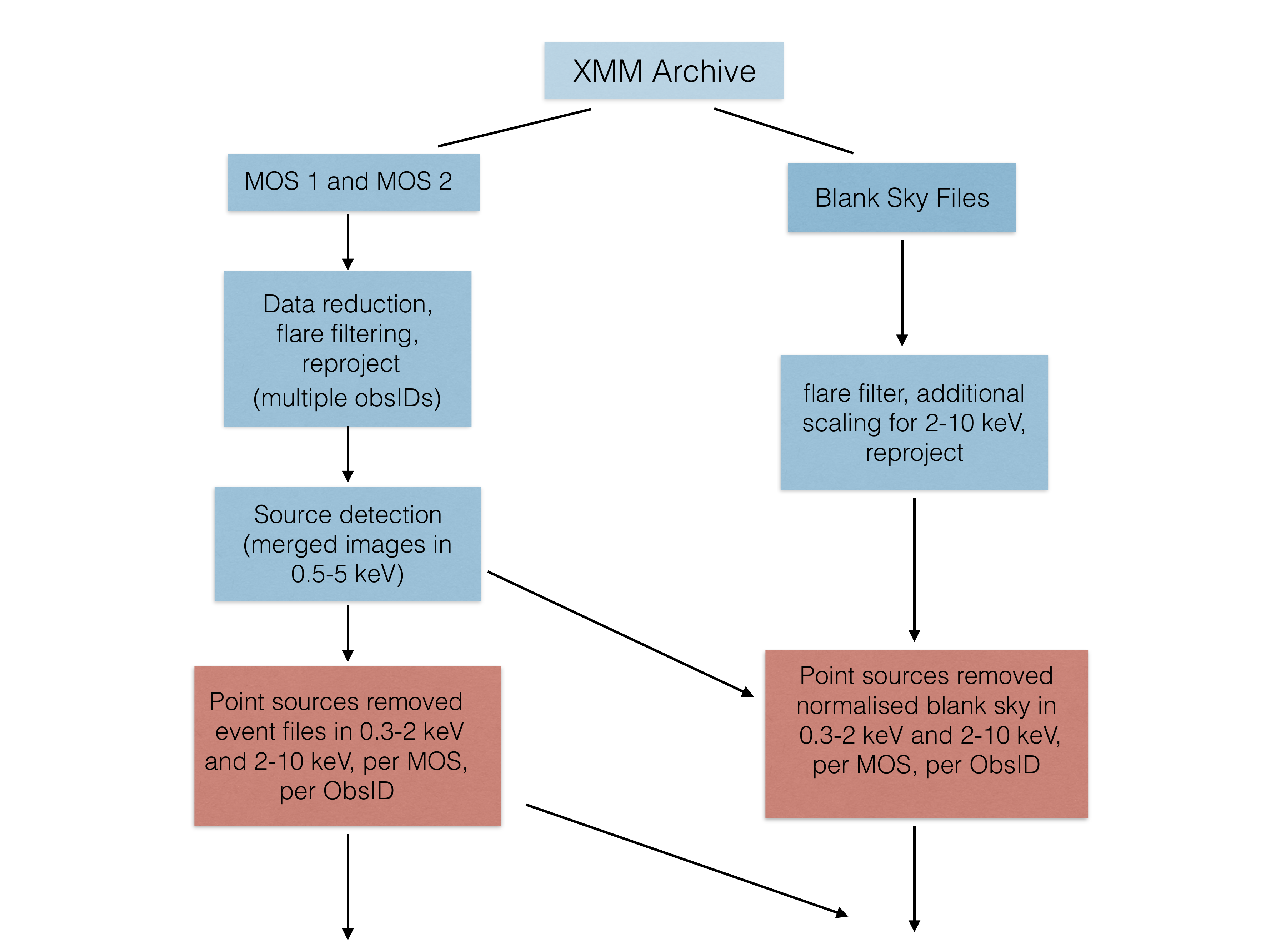}
 \includegraphics[scale=0.4,angle=0]{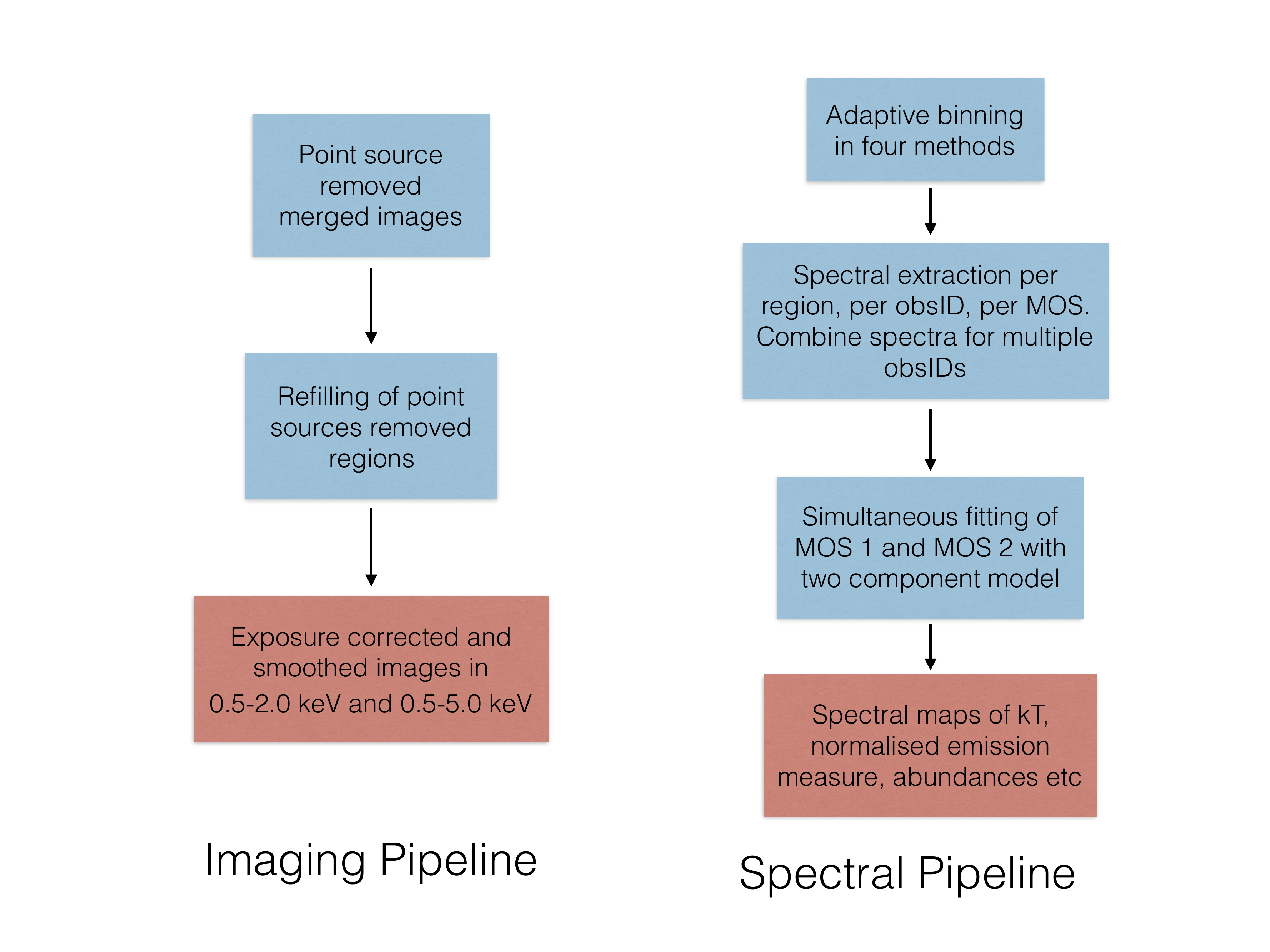}
\caption{Overview of the data analysis flowchart. The top plot shows the data reduction steps done to the event and blank sky files. The lower left plot shows the imaging pipeline, used in constructing the diffuse gas emission images. The lower right plot shows the spectral pipeline, used in constructing the spectral maps. The details of the various steps in these pipelines are mentioned in detail in Section 3.}
\label{flowchart}
\end{figure*}

\section{Data Analysis} \label{data}
We have developed a robust data analysis pipeline, modeled on the methods developed for the CGA by K19, using SAS threads\footnote{https://www.cosmos.esa.int/web/xmm-newton/sas-threads}. Figure \ref{flowchart} shows flowcharts of the process.

The main steps in the data reduction and analysis are:

\begin{enumerate}
\item Filtering of the MOS event files, removing background flares, applying up-to-date calibration and tailoring blank files for background estimation (Section 3.1.1). 
\item Carrying out source detection on the merged image and removing detected sources to allow clean imaging of the diffuse emission (Section 3.1.2).
\item Adaptively binning with four different binning methods, to determine the optimal spectral extraction regions (Section 3.2).
\item Extracting spectra, fitting them with suitable models and mapping the spectral parameters (Section 3.3 and 3.4).
\end{enumerate}

\subsection{Reduction of XMM-Newton data}
For a given galaxy, the MOS 1 and MOS 2 data for all the observations are downloaded from XMM Science Archive (XSA) and processed with SAS version 16.1.0. The SAS tool {\tt emchain} was used to re-reduce the observations and generate the event files with the latest calibration. The correct blank sky files, corresponding to the observation mode and filter are also downloaded.\footnote{https://xmm-tools.cosmos.esa.int/external/xmm$\_$calibration/background/bs$\_$repository/blanksky$\_$all.html}. For a galaxy having multiple observations, all the observations were reprojected to the tangent plane of the first observation. We first filter the event lists using standard filters for MOS ((PATTERN $\leq$ 12) \&\& (PI in [200:12000]) \&\&\#XMMEA$\_$EM).

\subsubsection{Background estimation}

The background estimation in XMM is important for extended sources which fill up the entire FOV and therefore do not allow estimation of the local background. The total background emission in XMM/MOS consists of time variable or flaring component, mostly particle background and a relatively constant sky background, comprising of various components like cosmic hard X-ray background and soft Galactic emission, and instrumental background components such as fluorescence lines. 
\par 
To remove the contribution from the flaring particle background, the light-curves of MOS 1 and MOS 2 were estimated in 9.5--12 keV energy bands and the time intervals corresponding to count-rates greater than 2$\sigma$ from the average value were excluded. These light-curves in a hard energy band were visually inspected to further estimate the times when the background rate is changing gradually throughout the observations or flaring occurs during a significant fraction of an observation. These times were manually removed before re-estimating the good time intervals within 2 $\sigma$ from the average.
\par
The three different methods of background estimation:
\begin{itemize}
\item Simple Background Subtraction: This method follows from the prescription by \cite{nevalainen2005}, where the blank sky files, appropriate to the observation and filter, are used after normalising the hard energy band count-rate to that of the event files.
\item Double Background Subtraction: This method is proposed by \cite{arnaud2001}, to separate the cosmic X-ray background from the instrumental background by using vignetting-corrected event files. The background spectra are selected from both blank sky files and a source free region of the event files. Since the emission from the galaxies in our sample fills up the entire field of view of MOS, this method of background estimation is not applicable in our analysis.
\item Background Modelling: This method follows the procedure outlined in``Cookbook for Analysis Procedures for XMM EPIC Observations of Extended Object and the Diffuse Background"\footnote{https://heasarc.gsfc.nasa.gov/docs/xmm/esas/cookbook/xmm-esas.html} \citep{snowden2011}. The background is partly subtracted and partly modelled in this method, hence it is complex and difficult to implement in the unsupervised fitting procedures developed as part of the pipeline.
\end{itemize} 

\cite{paggi2017} compared the above three methods of background estimation and found that the results were not heavily dependant on the method of background estimation. Hence we adopted the procedure for background estimation by \cite{nevalainen2005}.  We downloaded the appropriate blank sky files for a given observation and filter\footnote{https://xmm-tools.cosmos.esa.int/external/xmm$\_$calibration/background/bs$\_$repository/blanksky$\_$all.html} \citep{carter2007}. These blank sky files were also filtered using good time intervals estimated from their constituent event files. As outlined in \cite{nevalainen2005}, we scaled the count-rates in the blank sky files in 2-10 keV to match the count-rates of the event files in 2-10 keV. The count-rates of the blank sky files in 0.3-2 keV were kept unchanged. Although the soft background might be somewhat different, it would not seriously affect our data products in finding hot gas structures in the spectral maps, given that the three different background methods produce consistent results. 

\subsubsection{Construction of diffuse gas images}
For the purpose of detecting the point sources, the MOS 1 and MOS 2 event files, for multiple observations if present, were merged. Nearby galaxies were excluded from this merged image. 
We ran the CIAO tool {\tt wavdetect} (with scale = 2, 4, 8, 16 and sigthresh = 10$^{-6}$ ), with an appropriate psf for XMM MOS,  on the C band (0.5 -- 5 keV) image. {\tt wavdetect} might detect false sources near the galaxy center, where the hot gas emission peaks, or sometimes miss the detection of real sources. In such cases, we manually checked the detected point sources and re-adjusted them.  The size of each individual point source was manually checked as well and re-adjusted as required. We excised the detected point sources to create the point sources removed event files and images in 0.3--10.0 keV, 0.3--2.0 keV and 2--10 keV, both for observations and blank-sky files. The exposure of the blank sky files were scaled in 2--10 keV energy-band, matching the count-rate of source files in 9.5--12 keV. These point sources removed source and scaled blank sky event files in soft band 0.3--2 keV and hard band 2--10 keV, were used for further  analysis. 
\par
To create the diffuse gas images, we refilled the excised regions of point sources in the point sources removed images with values interpolated from the surrounding pixels using CIAO tools {\tt roi}, {\tt splitroi} and {\tt dmfilth}. We then generate the exposure-corrected and smoothed images of point-sourced removed and refilled with the CIAO tools {\tt aconvolve}, in 0.5--2.0 keV and 0.5--5.0 keV. We note here that merging of different observations was done only to run source detection, create diffuse gas images and adaptive binning (Section 3.2). For the purpose of extracting spectra, we used the point source removed event and scaled blank sky files in 0.3--2.0 keV and 2.0--10.0 keV, separately for different MOS detectors and different observations.

\subsection{Adaptive Binning}
 We have applied the following four adaptive binning methods, similar to those used in K19, to characterise the 2D spatial and spectral properties of the hot gas in the galaxies. 
 \begin{enumerate}
 \item Annulus Binning (AB): In this method, we use circular annuli, where the inner and outer radii of each annulus is adaptively determined based on S/N. This method has been widely used to study the 1D radial and spectral profiles of ETGs. The difficulty with this method is that we cannot infer the asymmetry in the distribution of gas in the galaxy since the regions are spherically symmetric.
 \item Weighted Voronoi Tessellation Binning (WVT or WB): Originally developed to analyse the optical integral field spectroscopic data by \cite{cappellari2003}, and later developed for X-ray data by \cite{diehl2006}, this method provides information on 2D maps of hot gas properties as well as underlying asymmetries in their distribution. 
 \item Contour Binning (CB): This method is similar to WB, where additionally it groups the areas with similar surface brightness. This utilises the method developed by \cite{sanders2006}. Similar to WB, this method provides 2D maps of hot gas properties as well as underlying asymmetries in their distributions. 
 \item Hybrid Binning (HB): This method utilises the grid-like binning method developed by \cite{osullivan2014}. Since this binning method results in overlapping extraction regions, the neighbouring bins are not statistically independent. However, it provides complementary information for regions with lower surface brightness with higher spatial resolutions, which is not possible with the other three binning methods. 
 \end{enumerate}
\par
All the above binning methods are applied to C band image (0.5--5.0 keV) and the size of the bin is determined by the requirement to achieve S/N = 50. While calculating the S/N for all the four binning methods, the background counts were considered for estimating the S/N $=$ src$\_$counts/(src$\_$counts + bg$\_$counts)$^{1/2}$

\subsection{Spectral extraction}
After performing the adaptive binning with four methods, we used the SAS tool {\tt evselect} to extract the X-ray spectra for each spatial bin. We use the point-source removed source and blank sky event files. For each bin, the X-ray spectra were extracted for each MOS detector, each ObsID, in soft (0.3--2.0 keV) and hard (2.0--10.0 keV) using the corresponding source and blank sky event files. The spectra extracted from the blank sky event files were scaled to match the count-rates of the event files as mentioned in Section 3.1.1. Hence for each spatial bin, two set of source and background spectral files were extracted, per ObsID, per MOS detector. We generated the response matrices (rmf) and ancillary response matrices (arf) using the SAS tools {\tt rmfgen} and {\tt arfgen}. We combined the spectral and rmf/arf files for different obsids, per bin, with the SAS task {\tt epicspeccombine}. The resulting combined spectra for each MOS and in soft and hard band were used in spectral fitting. We also performed simultaneous fit for the multiple ObsIDs without combining them and found no significant difference in the estimated spectral parameters.

\subsection{Spectral Fitting}
We carried out joint spectral fits to MOS 1 and MOS 2 spectral files, soft (0.3--2.0 keV) and hard band (2--10 keV). The spectral fitting was done with {\tt Sherpa}. The spectra were grouped for a minimum of 20 counts, to apply $\chi^{2}$ statistics. We fitted a two-component emission model {\tt VAPEC} \footnote{https://heasarc.gsfc.nasa.gov/xanadu/xspec/manual/node134.html} for hot gas (collisionally ionized diffuse gas) and a power-law with index 1.7 for undetected low mass X-ray binaries \citep{boroson2011}, to the spectral files. We carried out a joint fit to the MOS 1 and MOS 2 spectra and added a constant (cross-normalisation between MOS 1 and MOS 2) to the above model. The line of sight column density of hydrogen N$_{H}$ was fixed to the Galactic HI column density \citep{dickey1990}. We fit the emission model {\tt VAPEC} with two metal abundance ratios: for solar abundances at GRSA \citep{grevesse1998} and all the metal abundances tied to Fe abundances.
\par
Based on the spectral fits, we produce maps of the various model parameters such as gas temperature (kT), normalised emission measure (normalisation of the {\tt VAPEC} model divided by the area of the bin), abundances, reduced $\chi^{2}$ etc. We apply a mask at different confidence levels of the spectral parameter (10 \%, 20\% and 30\%) to show only the bins with values of the parameters having errors less than the confidence limit applied. We also produce projected pseudo-entropy (K$_{P} \sim$ S$_{X}^{1/3}$ T) and projected pseudo-pressure maps (P$_{P} \sim$ S$_{X}^{1/2}$ T), in arbitrary units, where S$_{X}$ is the normalised emission measure.
\par
Since the four adaptive binning methods create large numbers of spatial bins, of the order of 10,000 for some galaxies with deep observations, we utilise the Smithsonian Institution High Performance Cluster (SI/HPC)\footnote{https://confluence.si.edu/display/HPC/High+Performance+Computing}. It is a Beowulf cluster consisting of nearly 4000 CPU cores, distributed over 108 compute nodes and 24TB of total RAM. We optimally process the spectral pipeline, including adaptive binning, spectral extraction for all the region files and spectral fitting and creating maps of temperatures, emission measure etc, using SI/HPC. An example of the 2D spectral temperature and Fe maps using the contour and WVT binning methods are shown in Figure 3 and Figure 5 for NGC 4636 and NGC 1550.

\section{Discussion}
Taking advantage of the large effective area and the wide field of view of the XMM-Newton observatory, we use the NGA data products to investigate a number of important scientific questions. For example, the NGA data allow us to trace the full extent of the gas halos of ETGs and study the faint emission from the outskirts of galaxies. This is crucial for studying the interaction of hot gas with its surroundings. Furthermore, the higher S/N spectra obtained with the large effective area makes it feasible to investigate the 2D distribution of metal abundances in the hot halos. Along with the temperature map, we construct Fe abundance maps as part of the NGA project. Note that Fe maps were not produced in CGA (K19). The spatial variation of the Fe abundance has direct implications on the metal enrichment (by mass loss of evolved stars and SN explosion) and the transport histories by internal mechanisms (SN driven winds and AGN driven buoyant bubbles) and external mechanisms (sloshing and ram-pressure stripping) (e.g., see \cite{kim2012}). In this paper, we present a few highlights which best demonstrate the XMM-Newton capability by presenting the hot ISM characteristics in two galaxies, NGC 4636 and NGC 1550, with a particular emphasis on the importance of the 2D spectral maps. We will present results using the entire NGA data products in a separate paper.

\subsection{NGC 4636}
NGC 4636 is an excellent example of small-scale hot gas features, which are related to internal effects, and of a large-scale extended halo, which is related to the external effects. This hot gas-rich elliptical galaxy (at D = 14.7 Mpc) lies at 10$^{\circ}$ to the S from the Virgo cluster center and at the northern end of the Virgo South Extension (centered around NGC 4697). The hot halo in this galaxy has been extensively studied using data from various X-ray observatories - Einstein \citep{forman1985}, ASCA \citep{awaki1994}, ROSAT \citep{trinchieri1994}, Chandra \citep{jones2002, johansson2009} and XMM-Newton \citep{osullivan2005, finoguenov2006, ahoranta2016}. Previous investigators have
produced an intensity map and spectral maps (e.g., \citealt{finoguenov2006} with XMM data; \citealt{diehl2007} with Chandra data; \citealt{osullivan2005} with both data) and in general we find consistent results. Determining spectral maps (including Fe maps) by multiple binning methods in a uniform manner and examining both Chandra and XMM data help us understand the overall picture of the hot gas evolution.

\begin{figure*}
\centering
 \includegraphics[scale=0.5,angle=0]{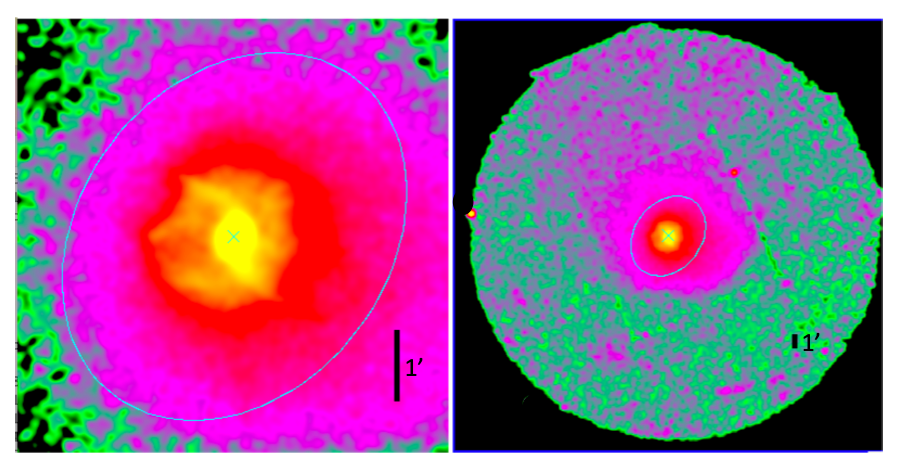}
\caption{The X-ray surface brightness maps (0.5-2 keV) of NGC 4636 made with the Chandra (left) and XMM (right) observations. The detected point sources were removed and filled with the photons from the surrounding area (see Section 3.1.2), and the exposure correction was applied. The cyan ellipse indicates the D$_{25}$ ellipse (semi-major axis = 3 arcmin or 13 kpc) in both images.}
\label{fig2}
\end{figure*}

\par
Figure \ref{fig2} shows the hot gas distribution of NGC 4636, after point sources removed and filled, produced with the Chandra (left) and  XMM (right) observations. In both panels, we overlay the D$_{25}$ ellipse (cyan) to indicate the stellar system's size and visualize different scales in two images. The Chandra observations reveal that on a small scale ($<$ 10 kpc, inside the D$_{25}$ ellipse), two cavities to the NE and SW from the center are surrounded by the spiral-arms-like features. The smaller-scale features are likely related to nuclear activities and radio jets (e.g., \citealt{jones2002, giacintucci2011}). Although fine details are not resolved, the XMM observations reveal asymmetrically distributed hot gas – (1) the small-scale ($\sim$ 10 kpc ) elongated structure to the NW-SE direction (similar to the major axis direction of the D25 ellipse, i.e., aligned to the stellar body) inside the D25, (2) the intermediate-scale ($\sim$ 20 kpc) extension toward the WSW direction beyond the D25 ellipse, and (3) the large-scale ($\sim$ 50 kpc) extension toward the N direction. The northern extension is faint but pronounced compared to the surface brightness at a similar distance toward the opposite S direction. See also \cite{finoguenov2006} and their Figure 4. The intermediate-scale and large-scale features (Fig 2 right) are the typical pattern of sloshing (e.g., \citealt{zuhone2016}) due to the perturbation by multiple passages of nearby galaxies (see also \citealt{osullivan2005, baldi2009}).

\begin{figure*}
\hspace{-30mm}
 \includegraphics[scale=0.8,angle=0]{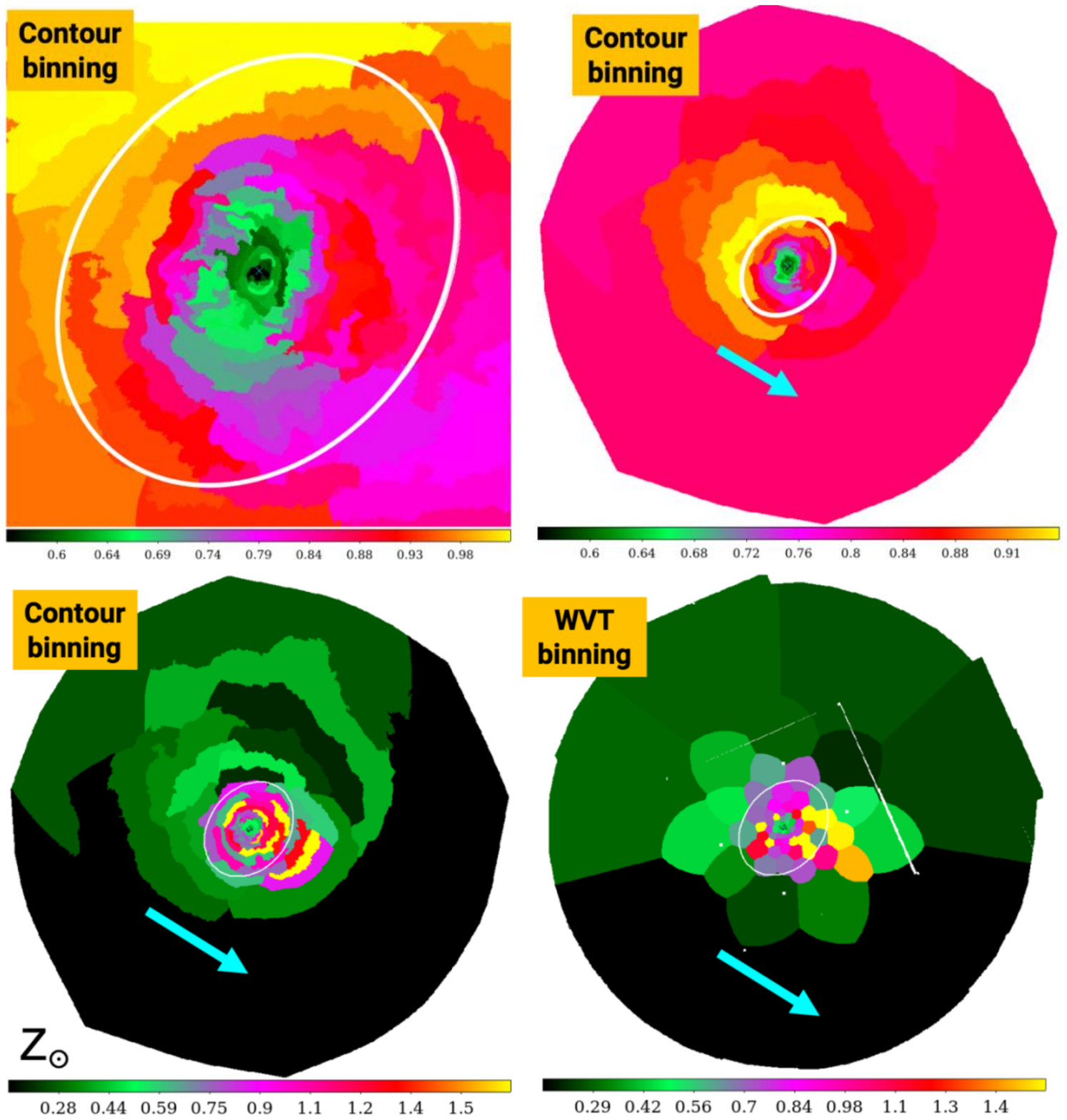}
\caption{Top panels. The temperature maps of NGC 4636 made with the Chandra (left) and XMM (right) observations. Bottom panels. The Fe abundance maps made with the XMM observations applying the contour binning (left) and the WVT binning (right). The ellipse indicates the D$_{25}$ ellipse (semi-major axis = 3 arcmin or 13 kpc) in all images. The arrow shows the direction of the Fe rich hot gas.}
\label{fig3}
\end{figure*}

\par
In Figure \ref{fig3}, we show the spectral maps of NGC 4636. The top panels show the T maps made with Chandra (left) and XMM (right) data. The T maps show the asymmetric distribution of the inner cooler gas of $\sim$0.6 keV, which is elongated to the NW-SE direction inside the D$_{25}$ ellipse, coincident with the small-scale elongation seen in the surface brightness map (Figure \ref{fig2}b). The gas temperature at the locations of the two cavities (in the Chandra T-map, also seen in Figure \ref{fig2}a) is slightly higher (about $\sim$0.8 keV) than that at the perpendicular direction at a similar distance ($\sim$0.6 keV), as expected by the pressure balance. 
\par
Toward the N and E directions from the center, the gas temperature increases with increasing radius and reaches its maximum kT $\sim$ 1 keV at r $\sim$ 15 kpc (just outside the D$_{25}$ ellipse). To the opposite direction (WSW – indicated by an arrow in Figure \ref{fig3} top-right), the gas temperature does not increase at the same rate. The intermediate-scale elongation to the WSW direction is cooler ($\sim$0.75 keV) than the gas in the opposite direction (1 keV). On a large scale, the northern extension is slightly cooler ($\sim$0.8 keV) than the opposite side ($\sim$0.9 keV). In summary, the gas in the small-scale elongation (NW-SW), intermediate-scale extension (to WSW), and large-scale enhancement (to N) seen in the surface brightness map (Figure \ref{fig2} right) is cooler than the gas at a similar distance to the opposite directions. 
\par
The Fe map further reveals interesting clues on the nature of the hot gas features. In the bottom panel of Figure \ref{fig3}, we show the Fe map constructed with the XMM observations in two adaptive binning methods (contour binning and WVT as described in Section 3.2).  Although the metal abundance and the radial variations in NGC 4636 were previously studied, the 2D Fe maps by multiple binning methods are presented here for the first time. In the central region ($<$ 3 kpc; green spatial bins), the Fe abundance is low ($\lesssim$ 0.5 solar). Apart from this central Fe-deficit, which we will discuss below, the hot gas is Fe-rich (about solar or higher; red and yellow spatial bins) inside the D$_{25}$ ellipse. Interestingly, the intermediate-scale cooler gas extended to the WSW elongation (seen in the SB map and T map) is richer in Fe ($\sim$1.5 x solar) than the hotter gas in the opposite direction at a similar distance. The large-scale northern extension is Fe-rich ($\sim$0.5 solar) compared to the gas at a similar distance to the opposite direction ($\sim$0.2 solar).
\par
In summary, the gas in the intermediate-scale extension to the WSW and the large-scale features (to N) seen in the surface brightness map is cooler and richer in Fe than the gas at a similar distance to the opposite directions. The projected pseudo-entropy map (K$_{P} \sim$  S$_{X} ^{-1/3}$T) shows the same trend because the dense, low T gas features have lower entropies (see the NGA web page\footnote{https://cxc.cfa.harvard.edu/GalaxyAtlas/NGA/v1}). We interpret this as an indication that the cooler, metal-enriched, low-entropy gas, originated from the stellar system (inside the D$_{25}$ ellipse) by mass loss and SN ejecta, is stretched out to the WSW on the intermediate scale (20 kpc) and the N on a large scale ($\sim$50 kpc) due to sloshing. The fact that extended gas directions are different with different radii is the typical phenomena of sloshing as the center of the galaxy has been perturbed, or sloshed, more than once (e.g., see the simulations by \citealt{zuhone2016}). 
\par
The Fe deficit in the very central region ($<$ 3 kpc) in NGC 4636 may require a different physical mechanism. The metallicity deficit at the center was reported in some gas-rich galaxies (e.g., \citealt{rasmussen2009, panagoulia2015}). However, given that the central region is the most complex in thermal and chemical structures of the multi-phase hot ISM, this measurement is challenging and may suffer from unknown systematic errors. Even with the two-temperature fit, which was often applied to account for the temperature gradient, the result may still be inconclusive, e.g., because of the limitation of modeling with two components and fixing elements in each component. If the Fe deficit in the center is real, which can be confirmed by the future mission (e.g., XRISM), the possible explanations include the resonance scattering, the He sedimentation, and the stellar and AGN feedback, which may play a role in reducing the observed Fe abundance (see the review in \citealt{kim2012}). \cite{panagoulia2015} showed that the central deficit is seen more often in the galaxy with an X-ray cavity and a shorter cooling time and suggested that Fe may be incorporated in the central dusty filaments, which are dragged outwards by the bubbling feedback process. 

\subsection{NGC 1550}
NGC 1550 is an exciting counter-example of NGC 4636. This group-dominant galaxy (at D=51.1 Mpc) belongs to one of the most X-ray luminous galaxies within 200 Mpc with L$_{X,GAS} \sim 10^{43}$ erg sec$^{-1}$ \citep{sun2003}. This galaxy has been extensively studied using the Chandra \citep{sun2003}, XMM \citep{kawaharada2009}, and Suzaku observations \citep{sato2010}. Although it does not meet the
condition ( $\Delta m_{12} > 2$) for a fossil group \citep{sun2003}, this system is close to a fossil-group as a relaxed, merger remnant \citep{jones2003, kawaharada2009, sato2010}.

\begin{figure*}
\centering
 \includegraphics[scale=0.5,angle=0]{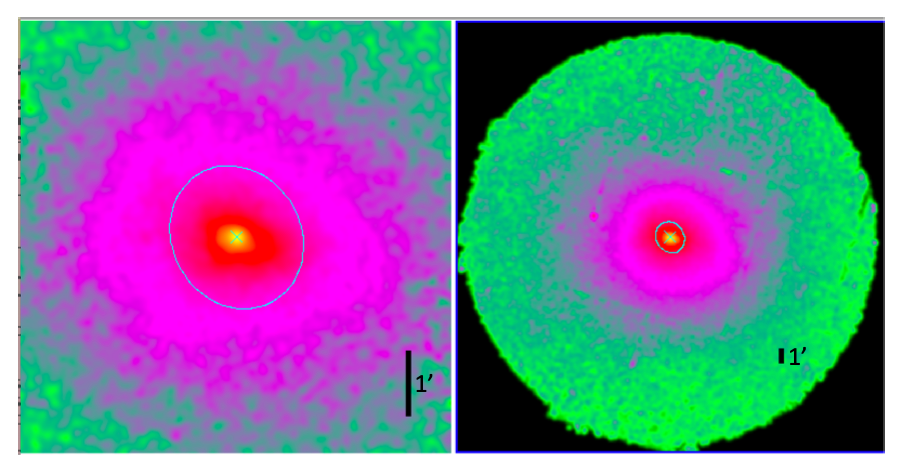}
\caption{The X-ray surface brightness maps (0.5-2 keV) of NGC 1550 made with the Chandra (left) and XMM (right) observations. The detected point sources were removed and filled with the photons from the surrounding area (see Section 3.1.2), and the exposure correction was applied. The cyan ellipse indicates the D$_{25}$ ellipse (semi-major axis = 1 arcmin or 17 kpc) in both images.}
\label{fig4}
\end{figure*}

\par
Unlike NGC 4636, the Chandra image (Figure \ref{fig4} left) indicates that the hot gas in NGC 1550 is relatively smooth, except in the central region where the Chandra image shows an E-W elongation (r $<$ 10 kpc; see also \citealt{sun2003, kolokythas2020}). The XMM observations (Figure \ref{fig4} right) show that on a large scale (50 - 100 kpc), the hot halo is smooth and relaxed (see also Figure 3 of \citealt{kawaharada2009}). In contrast to NGC 4636, this fossil-like system contains a relaxed hot halo, as expected as an end product of galaxy mergers (e.g., \citealt{jones2000, khosroshahi2006}). However, the spectral maps (temperature and abundance maps) exhibit interesting features (see below) that are not seen in the SB map and can provide essential clues on the nature of the hot halos.

\begin{figure*}
\vspace{-10mm}
\hspace{-30mm}
 \includegraphics[scale=0.8,angle=0]{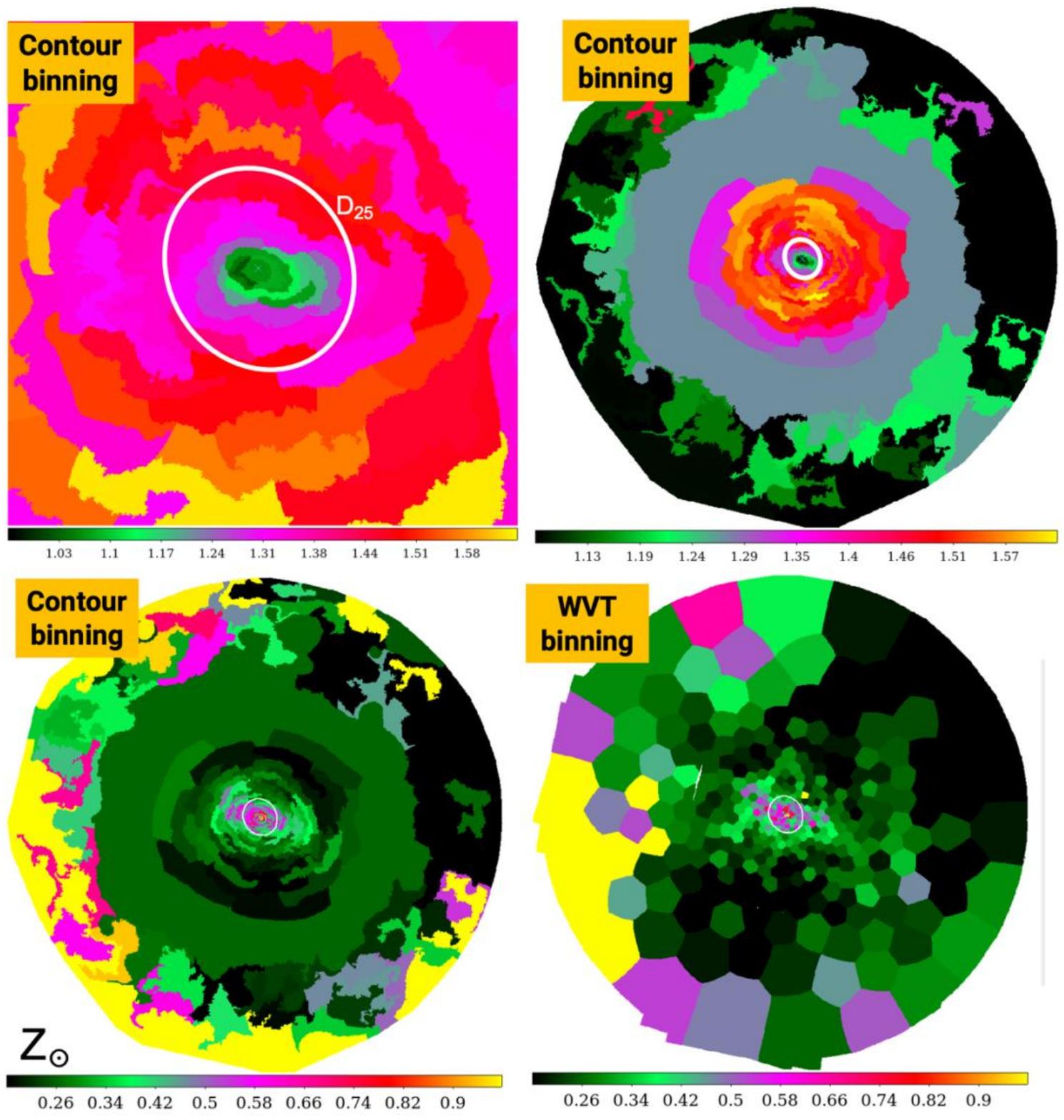}
\caption{Top panels. The temperature maps of NGC 1550  made with the Chandra (left) and XMM (right) observations. Bottom panels. The Fe abundance maps made with the XMM observations applying the contour binning (left) and the WVT binning (right). The ellipse indicates the D$_{25}$ ellipse (semi-major axis = 1 arcmin or 17 kpc) in all images.}
\label{fig5}
\end{figure*}

\par
In Figure \ref{fig5}, we show the spectral maps of NGC 1550. The top panels show the T maps made with Chandra (left) and XMM (right) data. On a small scale ($<$15 kpc), the 2D temperature map reveals the asymmetric distribution of cooler gas ($<$1 keV), which is elongated along the E-W direction, more pronounced to the W. Note that this is not aligned with the major axis of the D$_{25}$ ellipse (PA=30$^{\circ}$), i.e., misaligned to the stellar body. On a large scale, the temperature map is smooth, similar to that seen in the SB map. The gas temperature peaks ($\sim$1.6 keV) at r = 30-50 kpc and slowly declines outward. 
\par
The bottom panel of Figure \ref{fig5} shows the Fe map constructed with the XMM observations in two adaptive binning methods (contour and WVT as described in Section 3.2). Although the metal abundance was previously studied \citep{sun2003, kawaharada2009, sato2010}, the 2D Fe map in NGC 1550 is presented here for the first time. The presence of elongated cooler gas is confirmed in the Fe map. The hot gas in the inner E-W elongation is richer in Fe ($\sim$ 0.8 x solar) than the gas at a similar distance in other directions ($\lesssim$ 0.5 x solar). On a large scale, the Fe abundance decreases with increasing radius out to r $\sim$ 100 kpc. The high Fe abundance near the edge of the FOV is uncertain because of large errors in the Fe measurement, partly because two CCDs (top and bottom of MOS1) were not used in two out of three observations. 
\par
Again, we interpret that the cooler, metal-enriched, low-entropy gas, originated from the stellar system (inside the D$_{25}$ ellipse) by mass loss and SN ejecta, is propagating primarily to the EW direction.

\begin{figure*}
\vspace{-20mm}
\hspace{-20mm}
 \includegraphics[scale=0.7,angle=0]{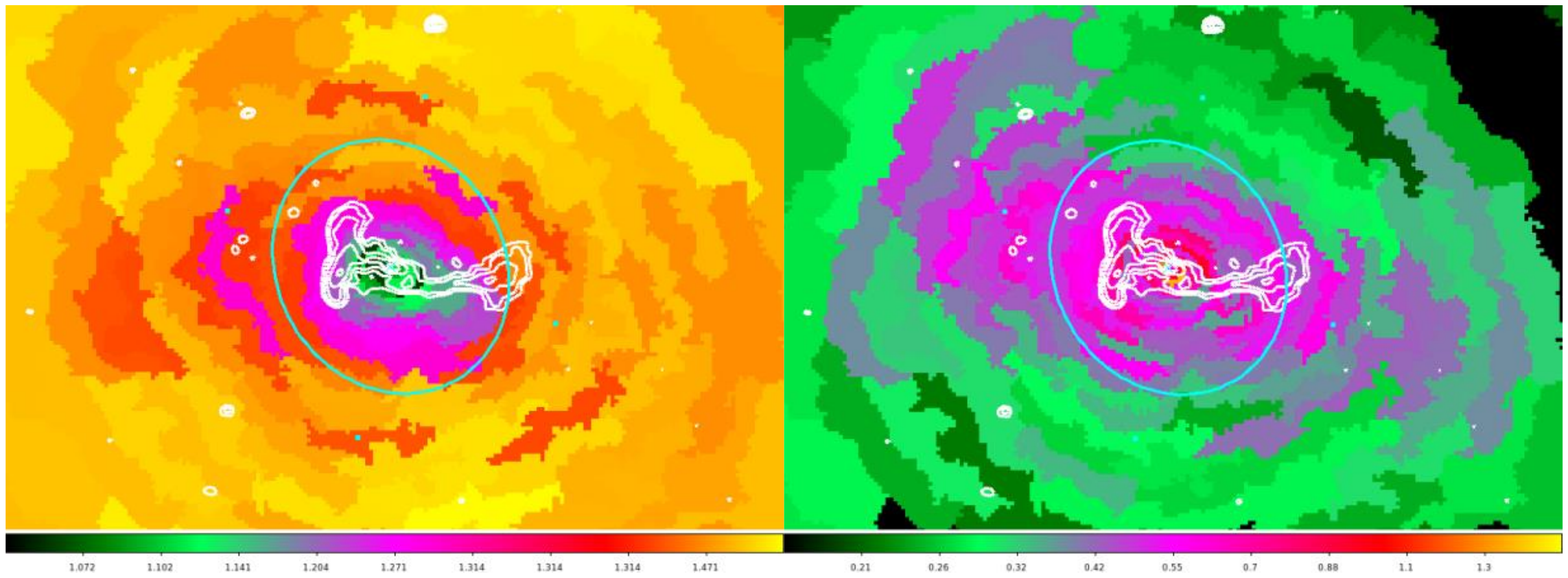}
 \vspace{-40mm}
\caption{GMRT 610 MHz radio contours (from \citealt{kolokythas2020}) showing the AGN jets and lobes are overlaid on the T map (left) and the Fe map (right). The cyan ellipse indicates the D$_{25}$ ellipse.}
\label{fig6}
\end{figure*}

\par
It is interesting to note that this E-W extension is aligned with the radio jet-lobe direction. The 1.4 GHz VLA radio observations of NGC 1550 \citep{dunn2010}. show two peaks: the primary peak close to the center and the secondary peak at $\sim$45'' to the W. The 610 and 235 MHz GMRT observations \citep{kolokythas2020} further reveal an asymmetric jet-lobe structure, aligned along the E-W direction, with a kink at the W jet and bending at the E jet before flaring up to the lobes in both sides. In Figure \ref{fig6}, the 610 MHz radio contours are overlaid on the T and Fe maps to show the alignment. \cite{kolokythas2020} examined the association between the radio jet-lobe structure and the asymmetric X-ray surface brightness feature. They considered the ideas of uplift and possible sloshing to explain the X-ray feature. 
\par
Here, we further show the apparent association between the T and Fe maps of the hot ISM and the radio structure, particularly to the West, indicating that the elongated cooler, metal-enriched, low-entropy gas is caused by radio lobes, likely by the uplift. The asymmetric Fe distribution, which is aligned with the radio jets and lobes, has previously been seen in a small number of clusters, the best example being Hydra A \citep{kirkpatrick2009}. See also \cite{mcnamara2016} for the theoretical consideration of a possible mechanism of lifting gas via AGN feedback. NGC 1550 represents a nearby, smaller-scale example of Hydra A. \cite{kirkpatrick2011} has determined the correlation between the jet power and Fe radius (R$_{Fe} \sim$ P$_{jet}^{0.42}$) in ten clusters with asymmetric Fe distributions (see their Figure 3). NGC 1550 roughly  follow this relation with P$_{jet}$(West) $\sim$  a few $10^{42}$ erg/s \citep{kolokythas2020}  and R$_{Fe} \sim$ 20 kpc (to the West) but falls at the lower-left corner in their Figure 3 with the lowest jet power and the smaller Fe radius. 
\par
On the contrary, the association of the E radio lode to the extension of the Fe-rich gas is visible
only inside the D$_{25}$ ellipse. As seen in Figure \ref{fig6}, the 2D Fe-rich, low-T gas distribution is asymmetric and extending far beyond the E lobe. While the E radio lobe stops at $\sim$10 kpc, the Fe-rich gas reaches to roughly twice the D$_{25}$ semi-major axis or at $\sim$40 kpc. Interestingly, this is the location of the possible sloshing front which \cite{kolokythas2020} identified in the residual image. Given that the extension of low-T, metal-enriched gas is better aligned with the
W radio lobe (than the E lobe) and that the E extension of low-T, metal-enriched gas lies far beyond the radio lobe and reaches the sloshing front (if real), we may be witnessing two mechanisms operating simultaneously - uplifting by the radio lobe to the W and sloshing to the E. \par
In both NGC 4636 and NGC 1550, the spectral maps reveal that the cooler, metal-enriched, low-entropy gas, originated from the stellar system, is propagating asymmetrically to the large distance from the center, likely due to interactions with AGN (uplift by radio jet-lobe) and with external galaxies (sloshing). To further illustrate the difference between the two galaxies, we compare the radial profiles of their spectral properties.

\begin{figure*}
\vspace{-20mm}
\hspace{-35mm}
 \includegraphics[width=24cm,height=20cm]{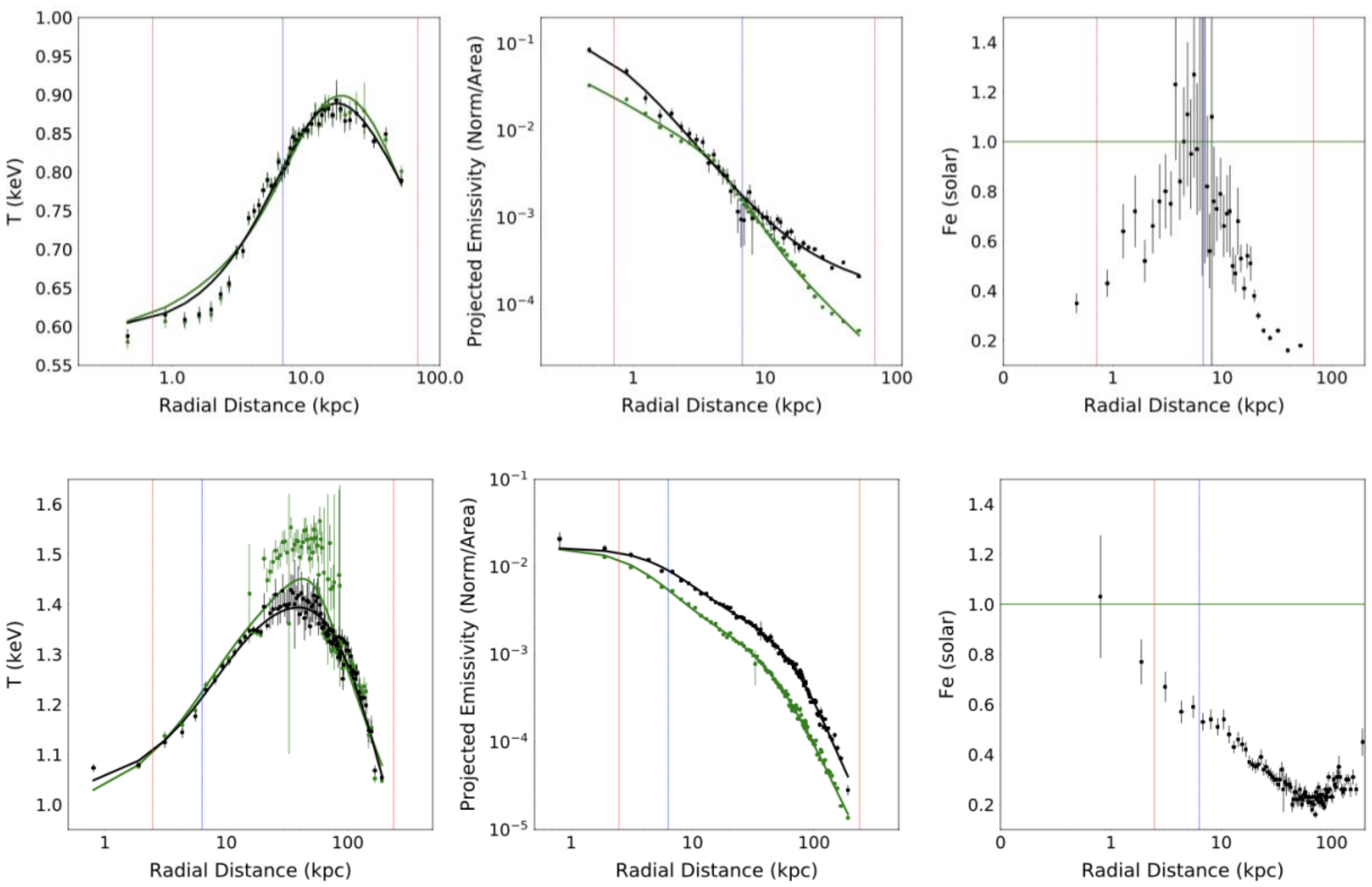}
 \vspace{-30mm}
\caption{The radial profiles of T (left), projected emissivity (middle), and Fe abundance (right) of the hot gas in (top panel) NGC 4636 and (bottom panel) NGC 1550. The spectral parameters are the azimuthally averaged values determined in the annulus binning. In all profiles, the black points are measured with the Fe abundance allowed to vary and the green points with the Fe abundance fixed at solar. The blue vertical line denotes the half-light radius, and the dotted red vertical lines are at r =10" and 16', which indicates the inner and outer boundaries set by the XMM psf and the field of view.}
\label{fig7}
\end{figure*}

\par
The radial profile of the hot gas temperature is indicative of various heating mechanisms like AGN feedback \citep{fabian2012}, stellar feedback \citep{ciotti1991}, and gravitational heating \citep{johansson2009}. For galaxy groups and clusters, there is a universal temperature profile, where the temperature is rising rapidly with increasing radius and peaks at r $\sim$ 0.1 R$_{VIR}$ (viral radius), and then slowly decreasing at larger radii \citep{vikhlinin2005, sun2009}. A similar trend in the temperature profiles is also seen in ETGs. \cite{kim2020} studied the radial profiles of 60 ETGs from CGA and propose a `universal' temperature profile of the hot halo in ETGs after considering various observational limitations and selection effects. With the XMM data, we can derive the radial profile of the Fe abundance.  
\par
Figure \ref{fig7} shows the radial profiles of the hot gas properties in NGC 4636 and NGC 1550. The azimuthally averaged quantities were determined with the annulus binning. The other binning methods produce similar results with larger scatters, which reflect the asymmetric distribution of the hot gas (i.e., different values at the same radius), as we described above. In all profiles, the black points are determined with the Fe abundance allowed to vary, and the green points are determined with the Fe abundance fixed at solar. The temperature profiles (left panels) in both galaxies belong to the hybrid-bump type (rising at small radii and falling at large radii) which is typical for a giant elliptical galaxy and the most common (40\%) profile among six types of ETGs \citep{kim2020}. 
\par
The Fe abundance profiles (right panels) in these two galaxies are quite different. The Fe abundance in NGC 4636 peaks at r $\sim$ 5 kpc and declines both to smaller and larger radii. The (peculiar) inward decrease makes the Fe deficit at the center, while the outward decline is expected because Fe is preferentially formed and released from the stars. On the other hand, the Fe abundance in NGC 1550 peaks at the center and monotonically declines with increasing radius. The apparent increase at r $>$ 100 kpc is likely unreal because of large systematic/statistical errors. 
\par
As described in \citealt{kim2020}, the T profiles of ETGs can be classified into six types, with a hybrid-bump type being the most common (43\%). The T profiles of NGC 1550 and NGC 4636 belong to this type (the left panel in Figure \ref{fig7}). Because the T profile is closely related to the cooling and heating mechanisms, the proper classification brought new insight into the hot gas thermal history (see \citealt{kim2020}). In the same terminology, the Fe profile of NGC 4636 is a hybrid-bump type, and that of NGC 1550 is a negative type. The complete classification of Fe profiles, which has not been thoroughly investigated, can allow us to address the chemical enrichment history and metal propagation in relation to the major evolutionary mechanisms, e.g., AGN/stellar feedback and environmental effect, as already seen above. We also note that the radial variation of the Fe abundance is reflected in the emissivity profile (middle panels) – the emissivity in NGC 4636 increases considerably both inside and outside where the Fe abundance is lower than solar and the emissivity in NGC 1550 increases in all radii except at the center. The variation in emissivity, in turn, affects the density-driven quantities, e.g., pressure and entropy. The full description of the Fe profile and its implication is beyond the scope of this paper. In a forthcoming paper, we will present the Fe profile classification and its related effect on hot gaseous halos.
\par
Our results are generally consistent with previously reported results whenever available. We note that the 2D Fe abundance maps  were rarely produced before. However, we have generated an extensive set of data products for each galaxy. Moreover, the spectral maps based on the four different spatial binning methods are complementary with each other. For example, AB shows the radial profiles, WB and CB show the statistically substantial quantities with associated errors, and HB can pick up small-scale features.  As shown in our two test cases (NGC 1550 and NGC 4636), with the NGA data products, we can address a complete picture in conjunction with the CGA products, including fine details and large-scale extended features.

\section{Summary}
We have developed robust pipelines to uniformly analyze the XMM data. Taking advantage of the high sensitivity and large field of view of the XMM, we have produced 2D spectral maps (e.g., temperature, emission measure, and Fe abundance) of hot gaseous halos in 38 early-type galaxies and made the data products publicly available at the dedicated website\footnote{https://cxc.cfa.harvard.edu/GalaxyAtlas/NGA/v1}.
\par
Our data products are most valuable to investigate the large-scale features of hot gaseous halos in ETGs and the distribution of metal enrichment throughout the galaxies, particularly when used in conjunction with the Chandra Galaxy Atlas (K19), which is optimal for examining the central region at the high spatial resolution. To illustrate the data quality and application, we describe two distinct galaxies – NC 4636 and NGC 1550 in detail. The spectral maps reveal that in both galaxies, the low-temperature, metal-enriched, low-entropy gas is propagating asymmetrically to large distances from the center, due to the internal (uplift by radio jet-lobe) and external (sloshing by nearby galaxies) mechanisms. In particular, it is interesting to note that both mechanisms are probably taking place in NGC 1550. The Fe radial profiles of the two galaxies are quite different (hybrid-bump in NGC 4636 and negative in NGC 1550), indicating that the Fe profile varies from one galaxy to another and could affect density-driven quantities of hot halos. We plan to address the full detail of the Fe profiles and the related implications in a separate paper.

\section{Acknowledgments}
We thank the anonymous referee for helpful comments.
This work was supported by Smithsonian 2018 Scholarly Study Program and by NASA contract NAS8–03060 (CXC).  The computations in this paper were conducted on the Smithsonian High Performance Cluster (SI/HPC), and the data analysis was supported by the CXC CIAO software. We have used the NASA ADS facilities. 

\software{CIAO (v4.10; \citealt{fruscione2006}), SAS (16.1.0; \citealt{gabriel2004}), Sherpa \citep{freeman2001}}

\section*{Appendix A: NGA Data Products}

The quick-look of the optical image, diffuse X-ray gas images in different energy bands, temperature, emission measure, pseudo pressure, pseudo entropy and Fe abundances maps constructed with the four binning methods are available on the NGA website. The following downloadable data products are made available in there as well. They include FITS files and ds9 PNG figures, which can be used for further analyses and directly compared with other wavelength data. The spectral maps (24 or 28) per (4) binning method per (6 or 7) spectral parameter are stored in two subfolders, one for fixed Fe and another for variable Fe. 
Since the number of spatial bins are large, the extracted source and background spectra along with its rmfs/arfs are not included in the data release. However, the FITS file `binno.fits' can be used to identify the bin number of the region of interest and the ASCII file `sum$\_$rad.dat' contains a summary of fitting results.

\begin{itemize}
\item \$\{gmv\}$\_$evt.fits.gz - a merged event file which combined-s all ObsIDs and both MOS1 and MOS2. \$\{gmv\} indicate galaxy name, merge id (mid) and version. \$\{gmv\} looks like N4636$\_$90101$\_$v01. 

\item \$\{gmv\}$\_$\{e\}$\_$img.fits - a combined image file. \$\{e\} is the energy band used, G 0.5-2.0 keV and C 0.5-5 keV.

\item \$\{gmv\}$\_$\{e\}$\_$exp.fits - a combined exposure map.

\item \$\{gmv\}$\_$\{e\}$\_$diff$\_$img.fits - a diffuse image file after subtracting and filling detected point sources.

\item \$\{gmv\}$\_$\{e\}$\_$sm$\_$diff$\_$flux.fits - a exposure-corrected, smoothed, diffuse image file.

\item \$\{gmv\}$\_$\{e\}$\_$sm$\_$diff$\_$flux.png  - same, but a ds9 png file

\item \$\{gmv\}$\_$src$\_$psfsize.fits - a FITS list of detected sources 

\item \$\{gmv\}$\_$OX.png  - a png file with an optical DSS image and a raw X-ray image side-by-side.

\item \$\{gmv\}$\_$rgb.png - a false (three) color image. The energy bands used in rgb are 0.5–1.2, 1.2–2, and 2–7 keV, respectively.

\item \$\{gm\}$\_$\{xB\}$\_$\{Smap\}.fits - a FITS image per each binning method and per each spectral parameter. xB indicates a binning method (AB, WB, CB, or HB). Smap indicates a spectral map (Imap for intensity, Tmap for temperature, Cmap for $\chi^{2}$, N for normalization/area or EM, P for pressure, K for entropy, or Fe for Fe abundance).

\item \$\{gm\}$\_$\{xB\}$\_$\{Smap\}.png - the same, but a ds9 png file. 

\item \$\{gm\}$\_$\{xB\}$\_$sum$\_$rad.dat - an ASCII table containing (1) bin number, (2) galacto-centric distance in arcmin determined by photon weighted mean distance from the galaxy center, (3) area of the bin in pixel (one pixel is 2" x 2") (4-5) total and net counts, (6) reduced $\chi^{2}$, (7-9) best-fit T and its 1-sigma lower and upper bounds, (10) T error in percent, (11-13) APEC normalization parameter divided by bin area and its 1-sigma lower and upper bounds (14-16) best-fit Fe abundance and its 1-sigma lower and upper bounds.

\item \$\{gm\}$\_$\{xB\}$\_$binno.fits  a FITS image file with pixel value = bin number. 

\end{itemize}

\newpage

\begin{table*}
\centering
\caption{NGA galaxy list}
\label{XRBs}
\begin{tabular}{c c c c c c c c c c c c c}
\footnote{Note: Column 1. Galaxy name (NGC or IC). Column 2--3: RA and DEC (J2000) from 2MASS via NED(http://ned.ipac.caltech.edu/). Column 4: Distance in Mpc. Column 5: Type taken from RC3 catalogue. Columns 6--7: Semi-major and semi-minor axis of the D$_{25}$ ellipse in arcminutes taken from RC3 catalogue. Column 8: Position angle of the D$_{25}$ ellipse from 2MASS via NED, measured eastward from the north. Column 9: Effective radius in arcmin taken from RC3 catalogue. Column 10: K-band luminosity from 2MASS (K$\_$tot mag) via NED (assuming M$_{K}(Sun)$ = 3.28 mag and D in Column 4). Column 11: Galactic line-of-sight column density of hydrogen in units of 10$^{20}$ cm$^{-2}$. Column 12: XMM-Newton merge id (mid -- see Table 2 for individual ObsIDs). Column 13: The total (MOS1 + MOS2) effective exposure in kilosec after background filtering (see Table 2 for effective exposures of individual ObsIDs)}\\
\hline
Name   & RA & DEC & D  & Type  & r$_{maj}$  &  r$_{min}$  &  PA  &  Re  & log(L$_{K}$)  &  N$_{H}$  &  mid & eff exp \\
\hline
I1262  & 17 33 2.0 & +43 45 34.6 & 130.0  & -5.0  & 0.60  & 0.32  & 80.0  & 0.20  & 11.42  & 2.43  & 90201 & 24.0 \\
I1459 & 22 57 10.6 & -36 27 44.0 & 29.2 & -5.0 & 2.62 & 1.90 & 42.5 & 0.62 & 11.54 & 1.17 & 90201 & 261 \\
I1860 & 02 49 33.7 & -31 11 21.0 & 93.8 & -5.0 & 0.87 & 0.60 & 6.4 & 0.31 & 11.57 & 2.05 & 90101 & 69.0 \\
I4296 & 13 36 39.0 & -33 57 57.2 & 50.8 & -5.0 & 1.69 & 1.62 & 45.0 & 0.80 & 11.74 & 4.09 & 90101 & 91.0 \\
N0383 & 01 07 24.9 & +32 24 45.0 & 63.4 & -3.0 & 0.79 & 0.71 & 25.0 & 0.34 & 11.54 & 5.41 & 90201 & 75.0 \\
N0499 & 01 23 11.5 & +33 27 38.0 & 54.5 & -2.5 & 0.81 & 0.64 & 70.0 & 0.28 & 11.31 & 5.21 & 90101 & 85.0 \\
N0507 & 01 23 40.0 & +33 15 20.0 & 63.8 & -2.0 & 1.55 & 1.55 & 60.0 & 0.69 & 11.62 & 5.23 & 90101 & 188.0 \\
N0533 & 01 25 31.4 & +01 45 32.8 & 76.9 & -5.0 & 1.90 & 1.17 & 47.5 & 0.72 & 11.73 & 3.07 & 90201 & 87.0 \\
N0720 & 01 53 0.5  & -13 44 19.2 & 27.7 & -5.0 & 2.34 & 1.20 & 140.0 & 0.60 & 11.31 & 1.58 & 90202 & 216.0 \\
N0741 & 01 56 21.0 & +05 37 44.0 & 70.9 & -5.0 & 1.48 & 1.44 & 90.0 & 0.64 & 11.72 & 4.44 & 90201 & 122.0 \\
N1132 & 02 52 51.8 & -01 16 28.8 & 95.0 & -4.5 & 1.26 & 0.67 & 150.0 & 0.56 & 11.58 & 5.19 & 90101 & 46.0 \\
N1316 & 03 22 41.7 & -37 12 29.6 & 21.5 & -2.0 & 6.01 & 4.26 & 47.5 & 1.22 & 11.76 & 2.13 & 90201 & 266.0 \\
N1332 & 03 26 17.3 & -21 20 7.3  & 22.9 & -3.0 & 2.34 & 0.72 & 112.5 & 0.46 & 11.23 & 2.30 & 90101 & 115.0 \\
N1399 & 03 38 29.1 & -35 27 2.7  & 19.9 & -5.0 & 3.46 & 3.23 & 150.0 & 0.81 & 11.41 & 1.49 & 90201 & 175.0 \\
N1404 & 03 38 51.9 & -35 35 39.8 & 21.0 & -5.0 & 1.66 & 1.48 & 162.5 & 0.45 & 11.25 & 1.51 & 90201 & 311.0 \\
N1407 & 03 40 11.9 & -18 34 48.4 & 28.8 & -5.0 & 2.29 & 2.13 & 60.0 & 1.06 & 11.57 & 5.42 & 90101 & 80.0 \\
N1550 & 04 19 37.9 & +02 24 35.7 & 51.1 & -3.2 & 1.12 & 0.97 & 30.0 & 0.43 & 11.24 & 11.25 & 90301 & 339.0 \\
N1600 & 04 31 39.9 & -05 05 10.0 & 57.4 & -5.0 & 1.23 & 0.83 & 5.0 & 0.81 & 11.63 & 4.86 & 90201 & 139.0 \\
N2300 & 07 32 20.0 & +85 42 34.2 & 30.4 & -2.0 & 1.41 & 1.02 & 108.0 & 0.55 & 11.25 & 5.49 & 90101 & 103.0 \\
N3402 & 10 50 26.1 & -12 50 42.3 & 64.9 & -4.0 & 1.04 & 1.04 & 170.0 & 0.47 & 11.39 & 4.50 & 90101 & 45.0 \\
N3842 & 11 44 2.1  & +19 56 59.0 & 97.0 & -5.0 & 0.71 & 0.51 & 175.0 & 0.63 & 11.67 & 2.27 & 90101 & 48.0 \\
N3923 & 11 51 1.8  & -28 48 22.0 & 22.9 & -5.0 & 2.94 & 1.95 & 47.5 & 0.88 & 11.45 & 6.30 & 90201 & 270.0 \\
N4261 & 12 19 23.2 & +05 49 30.8 & 31.6 & -5.0 & 2.04 & 1.82 & 172.5 & 0.75 & 11.43 & 1.58 & 90201 & 226.0 \\
N4278 & 12 20 6.8  & +29 16 50.7 & 16.1 & -5.0 & 2.04 & 1.90 & 27.5 & 0.56 & 10.87 & 1.76 & 90101 & 57.0 \\
N4325 & 12 23 6.7  & +10 37 16.0 & 110.0 & 0.0 & 0.48 & 0.32 & 175.0 & 0.33 & 11.29 & 2.14 & 90101 & 41.0 \\
N4374 & 12 25 3.7  & +12 53 13.1 & 18.4 & -5.0 & 3.23 & 2.81 & 122.5 & 1.02 & 11.37 & 2.78 & 90101 & 101.0 \\
N4406 & 12 26 11.7 & +12 56 46.0 & 17.1 & -5.0 & 4.46 & 2.88 & 125.0 & 2.07 & 11.36 & 2.69 & 90101 & 138.0 \\
N4477 & 12 30 2.2  & +13 38 11.8 & 16.5 & -2.0 & 1.90 & 1.73 & 40.0 & 0.73 & 10.83 & 2.65 & 90101 & 27.0 \\
N4552 & 12 35 39.8 & +12 33 22.8 & 15.3 & -5.0 & 2.56 & 2.34 & 150.0 & 0.68 & 11.01 & 2.56 & 90101 & 49.0 \\
N4594 & 12 39 59.4 & -11 37 23.0 & 9.8  & 1.0  & 4.35 & 1.77 & 87.5 & 1.19 & 11.33 & 3.67 & 90101 & 46.0 \\
N4636 & 12 42 49.9 & +02 41 16.0 & 14.7 & -5.0 & 3.01 & 2.34 & 142.5 & 1.56 & 11.10 & 1.82 & 90101 & 116.0 \\
N4649 & 12 43 40.0 & +11 33 9.7  & 16.8 & -5.0 & 3.71 & 3.01 & 107.5 & 1.28 & 11.49 & 2.13 & 90201 & 240.0 \\
N5044 & 13 15 24.0 & -16 23 7.9  & 31.2 & -5.0 & 1.48 & 1.48 & 10.0 & 0.42 & 11.24 & 4.94 & 90202 & 250.0 \\
N5813 & 15 01 11.3 & +01 42 7.1  & 32.2 & -5.0 & 2.08 & 1.51 & 130.0 & 0.89 & 11.38 & 4.25 & 90301 & 301.0 \\
N5846 & 15 06 29.3 & +01 36 20.2 & 24.9 & -5.0 & 2.04 & 1.90 & 27.5 & 0.99 & 11.34 & 4.24 & 90301 & 346.0 \\
N6338 & 17 15 23.0 & +57 24 40.0 & 123.0& -2.0 & 0.76 & 0.51 & 15.0 & 0.48 & 11.75 & 2.60 & 90201 & 138.0 \\
N6482 & 17 51 48.8 & +23 04 19.0 & 58.4 & -5.0 & 1.00 & 0.85 & 65.0 & 0.37 & 11.52 & 7.77 & 90301 & 38.0 \\ 
N7618 & 23 19 47.2 & +42 51 9.5  & 74.0 & -5.0 & 0.60 & 0.50 & 10.0 & 0.36 & 11.46 & 11.93 & 90101 & 39.0 \\
N7619 & 23 20 14.5 & +08 12 22.5 & 53.0 & -5.0 & 1.26 & 1.15 & 40.0 & 0.57 & 11.57 & 5.04 & 90101 & 79.0 \\

\hline
\end{tabular}
\end{table*}

\newpage

\begin{center}
\begin{longtable}{|c|c|c|c|c|c|}
\caption{XMM-Newton observation log}\\
\hline
\multicolumn{1}{|c|}{\textbf{Name}} & \multicolumn{1}{c|}{\textbf{mid}} & \multicolumn{1}{c|}{\textbf{ObsID}} & \multicolumn{1}{c|}{\textbf{Obs$\_$Date}} & \multicolumn{1}{c|}{\textbf{OAA}} & \multicolumn{1}{c|}{\textbf{eff$\_$exp}} \\ \hline 
\endfirsthead

\multicolumn{6}{c}%
{{\bfseries \tablename\ \thetable{} -- continued from previous page}} \\
\hline \multicolumn{1}{|c|}{\textbf{Name}} & \multicolumn{1}{c|}{\textbf{mid}} & \multicolumn{1}{c|}{\textbf{ObsID}} &
\multicolumn{1}{c|}{\textbf{Obs$\_$Date}} &
\multicolumn{1}{c|}{\textbf{OAA}} &
\multicolumn{1}{c|}{\textbf{eff$\_$exp}} \\ \hline 
\endhead

I1262  & 90201 & 0021140901 & 2003-01-30 & 1.03 & 6.0  \\
       &       & 0741580201 & 2014-11-07 & 0.26 & 18.0 \\
I1459  & 90201 & 0135980201 & 2002-04-30 & 0 & 57.0 \\
       &       & 0760870101 & 2015-11-02 & 0 & 204.0 \\
I1860  & 90101 & 0146510401 & 2003-02-04 & 0.04 & 69.0 \\
I4296  & 90101 & 0672870101 & 2011-07-11 & 2.7 & 91.0 \\
N0383  & 90201 & 0305290101 & 2005-08-03 & 0.1 & 13.0 \\
       &       & 0551720101 & 2008-07-01 & 0   & 62.0 \\
N0499  & 90101 & 0501280101 & 2007-08-15 & 2.29 & 85.0 \\
N0507  & 90101 & 0723800301 & 2013-07-14 & 0.03 & 188.0 \\
N0533  & 90201 & 0109860101 & 2000-12-31 & 0.46 & 75.0 \\
       &       & 0109860201 & 2001-01-01 & 0.46 & 12.0 \\
N0720  & 90202 & 0112300101 & 2002-07-13 & 0.15 & 47.0 \\
       &       & 0602010101 & 2009-12-23 & 0.04 & 169.0 \\
N0741  & 90201 & 0153030701 & 2004-01-03 & 0.68 & 16.0 \\
       &       & 0748190101 & 2015-01-08 & 1.73 & 106.0\\
N1132  & 90101 & 0151490101 & 2003-07-16 & 0.03 & 46.0 \\  
N1316  & 90201 & 0302780101 & 2005-08-11 & 0.05 & 168.0 \\
       &       & 0502070201 & 2007-08-19 & 0.05 & 98.0 \\ 
N1332  & 90101 & 0304190101 & 2006-01-15 & 0    & 115.0 \\
N1399  & 90201 & 0012830101 & 2001-06-27 & 0.05 & 12.0 \\
       &       & 0400620101 & 2006-08-23 & 0.01 & 163.0 \\
N1404  & 90201 & 0304940101 & 2005-07-30 & 0    & 56.0 \\
       &       & 0781350101 & 2016-12-29 & 0    & 155.0 \\
N1407  & 90101 & 0404750101 & 2007-02-11 & 8.67 & 80.0 \\
N1550  & 90301 & 0152150101 & 2003-02-22 & 0    & 45.0 \\
       &       & 0723800401 & 2014-02-02 & 0.01 & 117.0 \\
       &       & 0723800501 & 2014-02-06 & 0.01 & 177.0 \\
N1600  & 90201 & 0400490101 & 2006-08-14 & 0.01 & 32.0 \\
       &       & 0400490201 & 2007-02-06 & 0.01 & 107.0 \\
N2300  & 90101 & 0022340201 & 2001-03-16 & 0.02 & 103.0 \\
N3402  & 90101 & 0146510301 & 2002-12-20 & 0.01 & 45.0 \\
N3842  & 90101 & 0602200101 & 2009-05-27 & 0.74 & 48.0 \\
N3923  & 90201 & 0027340101 & 2002-01-03 & 0.08 & 73.0 \\
       &       & 0602010301 & 2009-06-23 & 0.07 & 197.0 \\
N4261  & 90201 & 0056340101 & 2001-12-16 & 0.02 & 56.0 \\
       &       & 0502120101 & 2007-12-16 & 0.05 & 170.0 \\
N4278  & 90101 & 0205010101 & 2004-05-23 & 0.01 & 57.0 \\   
N4325  & 90101 & 0108860101 & 2000-12-24 & 0.11 & 41.0 \\
N4374  & 90101 & 0673310101 & 2011-06-01 & 0.02 & 101.0 \\
N4406  & 90101 & 0108260201 & 2002-07-01 & 0.06 & 138.0 \\
N4472  & 90301 & 0761630101 & 2016-01-05 & 6.71 & 190.0 \\
       &       & 0761630201 & 2016-01-07 & 6.71 & 177.0 \\
       &       & 0761630301 & 2016-01-09 & 6.71 & 174.0 \\
N4552  & 90101 & 0141570101 & 2003-07-10 & 0.01 & 49.0 \\
N4594  & 90101 & 0084030101 & 2001-12-28 & 0.01 & 46.0 \\
N4636  & 90101 & 0111190701 & 2001-01-05 & 0.47 & 116.0 \\
N4649  & 90201 & 0021540201 & 2001-01-02 & 0.06 & 96.0 \\
       &       & 0502160101 & 2007-12-19 & 0.1  & 144.0 \\
N5044  & 90202 & 0037950101 & 2001-01-12 & 0.03 & 42.0 \\
       &       & 0554680101 & 2008-12-27 & 0.03 & 208.0 \\
N5813  & 90301 & 0302460101 & 2005-07-23 & 0.02 & 58.0 \\
       &       & 0554680201 & 2009-02-11 & 0.02 & 122.0 \\
       &       & 0554680301 & 2009-02-17 & 0.02 & 121.0 \\
N5846  & 90301 & 0021540501 & 2001-08-26 & 0.01 & 29.0 \\
       &       & 0723800101 & 2014-01-21 & 0.05 & 149.0 \\
       &       & 0723800201 & 2014-01-17 & 0.05 & 168.0 \\
N6338  & 90201 & 0741580101 & 2014-12-04 & 0.26 & 24.0 \\
       &       & 0792790101 & 2016-10-12 & 0.66 & 114.0 \\
N6482  & 90301 & 0304160401 & 2006-02-18 & 0    & 17.0 \\
       &       & 0304160501 & 2006-03-20 & 0    & 3.0 \\
       &       & 0304160801 & 2006-04-13 & 0    & 18.0 \\
N7618  & 90101 & 0302320201 & 2006-01-20 & 6.11 & 39.0 \\
N7619  & 90101 & 0149240101 & 2003-12-16 & 2.12 & 79.0 \\

\hline
\end{longtable}
\footnotesize{Column 1: Galaxy name (NGC or IC). Column 2: Merge id (or mid) is an identification number denoting the number of ObsIDs used for analysis. Column 3: Unique XMM-Newton observation ID. Column 4: Date of observation. Column 5: Off axis angle (OAA) of the galaxy center in arcminutes. Column 6: Effective exposures of individual ObsIDs in kilosec after background filtering.}
\end{center}

\end{document}